\documentclass{aa}
\usepackage{graphicx}
\usepackage{natbib}
\usepackage{txfonts}
\bibpunct{(}{)}{;}{a}{}{,}    

\begin{document}
\title{NLTE solar irradiance modeling with the COSI code}
\author{A. I. Shapiro \inst{1}  \and W. Schmutz \inst{1} \and M. Schoell \inst{1,2} \and M. Haberreiter \inst{3} \and E. Rozanov \inst{1,4}}
\offprints{A.I. Shapiro}

\institute{Physikalisch-Meteorologishes Observatorium Davos, World Radiation Center, 7260 Davos Dorf, Switzerland\\
\email{alexander.shapiro@pmodwrc.ch}
\and Institute for Astronomy ETH, Zurich, Switzerland
\and  Laboratory for Atmospheric and Space Physics, University of Colorado, Boulder, CO 80303, USA
\and Institute for Atmospheric and Climate science ETH, Zurich, Switzerland\\}
\date{Received 30 December 2009; accepted 29 March 2010}

\abstract 
{The solar irradiance is known to change on time scales of minutes to decades, and it is suspected that its substantial fluctuations are partially responsible for climate variations. }
{We are developing a solar atmosphere code that allows the  physical modeling of the entire solar spectrum composed of quiet Sun and active regions. This code is a tool for modeling the variability of the solar irradiance and understanding its influence on Earth. }
{We exploit further development of the radiative transfer code COSI that now incorporates the calculation of  molecular lines. We validated COSI under the conditions of  local thermodynamic equilibrium (LTE) against the synthetic spectra calculated with the ATLAS code. The synthetic solar spectra were also calculated in non-local thermodynamic equilibrium (NLTE) and compared to the available measured spectra. In doing so we have defined the main problems of the modeling, e.g., the lack of opacity in the UV part of the spectrum and the inconsistency in the calculations of the visible continuum level, and we describe a solution to these problems. }
{The improved version of COSI allows us to reach good agreement between the calculated and observed solar spectra as measured by SOLSTICE and SIM onboard the SORCE satellite  and ATLAS 3 mission operated from the Space Shuttle.  We find that NLTE effects are very important for the modeling of the solar spectrum even in the visual part of the spectrum and for its variability over the entire solar spectrum. In addition to the strong effect on the UV part of the spectrum, NLTE effects influence the concentration of the negative ion of hydrogen, which results in a significant change of the visible continuum level and the irradiance variability.}
{}


\titlerunning{NLTE Solar Irradiance Modeling with the COSI code}

\maketitle
%
\section{Introduction}\label{sec:intro}
The solar radiation is the main source of the input of energy to the terrestrial atmosphere, so that it determines Earth's thermal balance and climate.  Although it has been known  since 1978 that the solar irradiance is not constant but instead varies on  scales from several minutes to decades \citep[cf.][]{froehlich2005,krivovasolanki2008}, the influence of this variability on the climate is not yet fully understood. 
Nowadays several datasets for the past spectral solar irradiance (SSI) based on the different reconstruction approaches \citep[e.g.][]{lean2005, krivovaetal2009} and satellite measurements are available.  However, the remaining disagreements between these data  lead to different atmospheric responses when they are used in the climate models  \citep{shapiroetal2009}. The task of constructing a self-consistent physical model in order to reconstruct the past solar spectral irradiance (SSI) remains of high importance.

Modern reconstructions of the SSI are based on the assumption that the irradiance changes are determined by the evolution of the solar surface magnetic field \citep[][]{foukallean1988, krivovaetal2003, domingoetal2009}. Areas of the solar disk are associated to several components (e.g. quiet Sun, bright network, plage, and sunspot) according to the measured surface magnetic field and the contrast  of the features. These components are represented by corresponding atmosphere structures \citep[cf.][]{kurucz1991,fontenlaetal1999}. The SSI is calculated by weighting the irradiance from each model with the corresponding filling factor.

The calculation of the emergent solar radiation, even from the atmosphere with known thermal structure, is a very sophisticated problem because consistent models have to account for the NLTE effects in the solar atmosphere. 
As the importance of these effects has become clear over the past several decades, several numerical codes have been developed. One of the first NLTE-codes, LINEAR, was published by \citet{aueretal1972}, who used a complete linearization method developed by \citet{auermihalas1969,auermihalas1970}.  Later, the MULTI code was published by \citet{carlsson1986}. This code is based on the linearization technique developed by \citet{scharmer1981} and  \citet{scharmercarlsson1985a, scharmercarlsson1985b}. More recently, the RH code,  which is based on the MALI (Multi-level Approximate Lambda Iteration) formalism of \citet{rybickihummer1991, rybickihummer1992} has been developed by \citet{uitenbroek2001}.

Combining the radiative transfer code  by  \citet{hamannschmutz1987,  schmutzetal1989} and  spectral synthesis code SYNSPEC by \citet{hubeny1981}, 
\citet{haberreiter2008} has developed the 1D spherical symmetrical COde for Solar Irradiance (COSI). The NLTE opacity distribution function (ODF) concept, which was implemented in this code, allows an indirect account for the NLTE effects in several million lines. This makes COSI especially suitable for calculating the overall energy distribution in the solar spectrum. In this paper we introduce a new version of COSI (version 2), describe the main modifications, and present first results.

In Sect.~\ref{sec:obs} we describe the datasets of the measured spectral irradiance, which were used for comparison with our calculated spectrum.
In Sect.~\ref{sec:molecules} we introduce the integration of molecular lines into COSI and show that it can solve the discrepancies to observations and to other codes. In Sect.~\ref{sec:NLTEproblems} we present the resulting solar spectrum from NLTE calculations. Due to missing opacity in the UV, the flux appears to be significantly higher than measured \citep{busaetal2001, shorthauschildt2009}. We solve this problem by introducing  additional opacity to the ODF for selected spectral ranges (Sect.~\ref{subsec:NLTEUV}),  while the problem of the NLTE visible continuum due to the deviations in the concentration of the hydrogen negative ion is addressed in Sect.~\ref{subsec:NLTEcont}. In Sect.~\ref{sec:active} we present the synthetic spectra of the active regions and its implications for the 
solar variability study. Finally, we summarize the main results in Sect.~\ref{sec:conclusions}.

\section{Measured solar spectra irradiance}\label{sec:obs}
We compared our calculated solar spectrum with several available datasets.
We used the observations taken with SOLSTICE (SOLar-STellar Irradiance Comparison Experiment,  \citet{SOLSTICE}) up to 320 nm and the Solar Irradiance Monitor (SIM) \citep{SIM} from 320 nm onward instruments onboard the SORCE satellite  \citep{rottman2005}  obtained during the 23rd solar cycle minimum (hereafter SORCE measurements). 
For the comparisons shown in this paper, we used the average of the observations from 21 April 2008 to 28 April 2008.

Our calculated spectrum was converted with a 1 nm boxcar profile for comparison with the SOLSTICE measurements and  trapezoidal profile    for comparison with the SIM measurements.   The SIM resolution strongly depends on wavelength (FWHM is about 1.5 nm for the 310 nm and about 22 nm for the 800 nm) so the parameters of this trapezoidal profile are also wavelength dependent and were provided by  \citet{harder2009}.

We also used the SOLar SPECtral Irradiance Measurements (SOLSPEC) \citep{thuilleretal2004} during the ATLAS 3 mission in November 1994.
For this comparison we convolved the calculated spectrum with a Gaussian with FWHM = 0.6 nm.

\section{Molecules in the COSI code}\label{sec:molecules}
COSI  simultaneously solves the equations of  statistical equilibrium and radiative transfer in 1D spherical geometry.  
All spectral lines in COSI  can be divided into two groups. The first group comprises about one thousand lines that are the most prominent in the solar spectrum. These lines are explicitly treated in NLTE. The second group contains several million background lines provided by \citet{kurucz2006} and calculated under the assumption that their upper and lower levels are populated in LTE relative to the explicit NLTE levels.  
These background lines are taken into account in the spectral synthesis part of the code but also affect NLTE calculations via the ODF. The ODF is iteratively recalculated until it becomes self-consistent with populations of the NLTE levels \citep{haberreiter2006, haberreiter2008}.  This allows us to indirectly account for the NLTE effects in the  background lines. 

Molecular lines were not included in the previous version of COSI \citep{haberreiter2008} . However, molecular lines play an important role in the formation of the solar spectrum.  In some spectral regions they are even the dominant opacity source. Therefore, a considerable amount of opacity was missing. This led to a discrepancy with observations. 
In this version of COSI  molecular lines were taken into account. This requires  computing  molecular concentrations, preparing the molecular line lists, and incorporating molecular opacities and emissivities into the code.

\subsection{Chemical equilibrium}\label{subsec:chem}
Under the assumption of the instantaneous chemical equilibrium, the concentration of molecule AB is connected with the concentrations of  atoms A and B  by the Guldberg-Waage law \citep[cf.][]{tatum1966}: 
\begin{equation}
\frac{n_{\rm A} n_{\rm B}}{n_{\rm AB}}=K_{\rm AB} (T)={ \left ( \frac{2 \pi m k T}{h^2} \right ) }^{3/2}  \frac{Q_{\rm A} Q_{\rm B}}{Q_{\rm AB}} \, kT \exp{(-D_0/kT)},
\label{eq:chemeq}
\end{equation}
where $K_{\rm AB}$ is the temperature dependent equilibrium constant, Q internal partition function for the atoms A and B, and molecule AB, $m$ is the reduced molecular mass, and $D_0$  the dissociation energy. 
The contribution to the opacity from the transitions between lower level $i$ and upper level $j$ of molecule AB is given by
\begin{equation}
\kappa_{\rm AB}^{ij}(\nu)=\frac{n_{\rm AB} \, g_i \exp{(-E_i/kT)}}{Q_{\rm AB}} \frac{h \nu}{4 \pi} \left (  n_i B_{ij} - n_j B_{ji} \right ) \phi(\nu - \nu_{ij}),
\label{eq:molopac}
\end{equation}
where $g_i$ is statistical weight of the lower level, B's are the Einstein coefficients, and $\phi(\nu)$ is the line profile normalized to one.

\begin{figure}
\resizebox{\hsize}{!}{\includegraphics{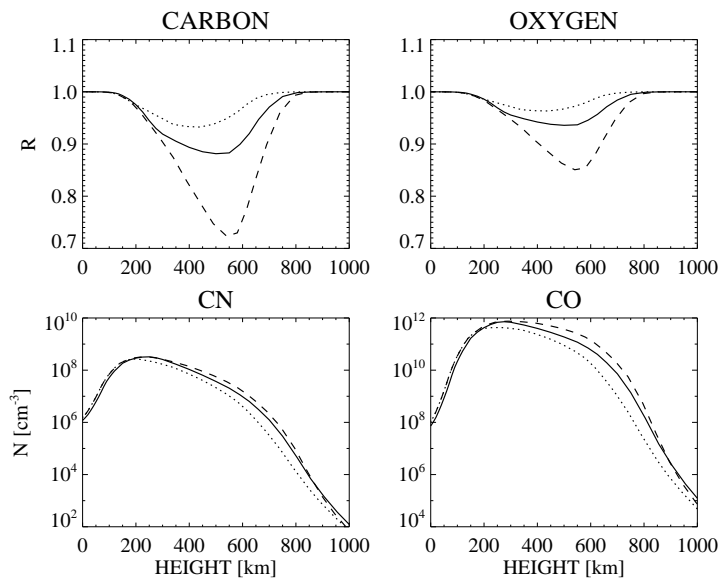}}
\caption{The ratio $R$ of the number concentration of carbon and oxygen that are not attached to the molecules to their total element amount (upper panels) and the CN and CO concentrations as a function of height (lower
panels) for three atmospheric models: FALA99 (dashed curves), FALC99 (continuous curves), and FALP99 (dotted curves).  The zero-height depth point in all models is defined as the radius at which the continuum optical depth at $5000 
\AA$ is equal to one. }
\label{fig:conc}
\end{figure}

There are several available temperature polynomial approximations  for the atomic and molecular partition functions, as well as for the equilibrium constants $K_{AB}$ \citep[cf.][]{tatum1966, tsuji1973,irwin1981, sauvalandtatum1984, rossietal1985}. Although molecular partition functions can contain significant errors due to the inaccurate values of the energy levels and even due to unknown electronic states \citep{irwin1981}, this does not significantly affect  the accuracy of molecular opacity calculations. This insensitivity exists because  the overall molecular concentration is proportional to the partition function (see Eq.~\ref{eq:chemeq}), while the opacity in a given line is inversely proportional to it (see Eq.~\ref{eq:molopac}). Therefore it is crucial to either calculate equilibrium constants directly from partition functions or use the same source for both approximations.

\citet{tsuji1973} shows that abundances of hydrogen, carbon, nitrogen, and oxygen can be determined quite accurately  without taking into account their dependencies on other elements. Because we are only interested  in molecular bands that can significantly contribute to the opacity over a broad spectral range, the system of equations like Eq.~(\ref{eq:chemeq}) were solved only for these four main elements and the  diatomic molecules built  from them. 

Because a significant fraction of atoms can be associated with molecules the chemical equilibrium calculation also affects the atomic lines. In Fig.~\ref{fig:conc}  we present the changes in the carbon and oxygen concentrations due to the association with 
molecules, together with CN and CO concentrations for three atmospheric models by \citet{fontenlaetal1999}:  the relatively cold supergranular cell center model FALA (hereafter FALA99), the averaged quiet Sun model FALC (hereafter FALC99), and the relatively warm bright network model FALP (hereafter FALP99).   Both molecular concentrations and deviations in atomic concentration show strong temperature sensitivities, and this is important for assessing of the solar spectral variability.

\subsection{Main molecular bands in the solar spectrum}\label{susec:bands}
The line list for the spectrum synthesis was compiled using the molecular databases of \citet{kurucz1993}  and Solar Radiation Physical Modeling (SRPM) database \citep[e.g.][]{fontenlaetal2009}, which is based on \citet{graycorbally1994} and HITRAN. Wavelengths and line strengths of the most significant lines of the CN violet system and CH G band  were calculated based on the molecular constants by \citet{krupp1974}, \citet{knowlesetal1988}, and \citet{wallaceetal1999}.  The OH and CH continuous opacities were calculated according to \citet{kuruczetal1987} under the LTE assumption.

\begin{figure}
\resizebox{\hsize}{!}{\includegraphics{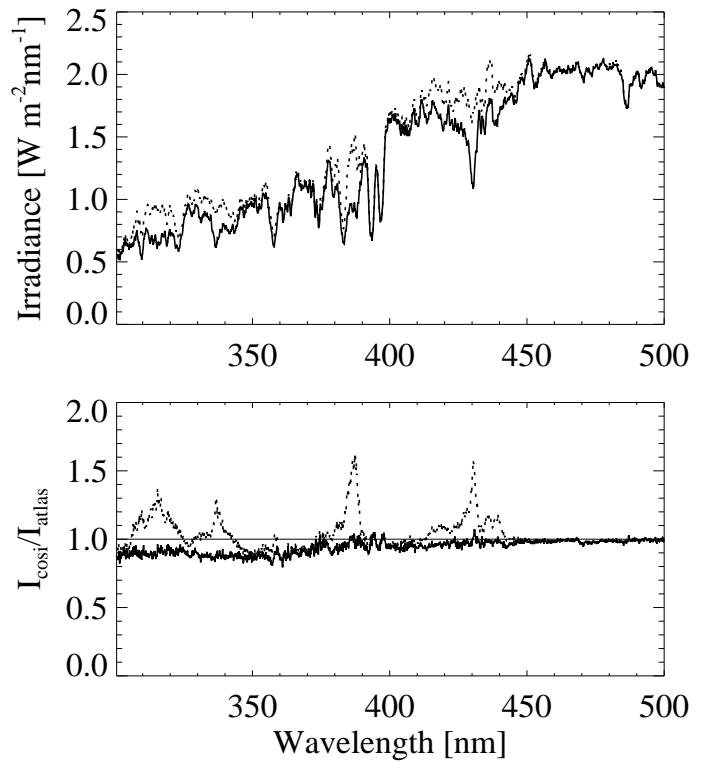}}
\caption{LTE calculation of molecular lines with COSI. Upper panel: Synthetic solar spectrum calculated with (solid line) and without (dashed line) molecular lines. Lower panel: Corresponding ratios to the irradiance calculated by the ATLAS12 code.  All spectra are averaged with a 1-nm boxcar.}
\label{fig:LTEmol}
\end{figure}

In Fig.~\ref{fig:LTEmol} we present the part of the solar spectrum where the LTE calculations with the previous version of COSI, which did not account for molecular lines, showed significant deviations from the calculations with the LTE radiative transfer code ATLAS12  carried out by \citet{kurucz2005}.  Both calculations used the same atmosphere structure and abundances by \citet{kurucz1991} (hereafter K91). One can see that introducing the molecular lines into COSI quite strongly affects the spectrum and significantly diminishes disagreements with the ATLAS 12 code. The most prominent features are the CH G band around 430 nm, CN violet G band around 380 nm, CN, NH, and OH bands between 300 and 350 nm. The remaining differences can possibly be attributed to the different photoionization cross sections implemented in COSI and ATLAS12.

\begin{figure}
\resizebox{\hsize}{!}{\includegraphics{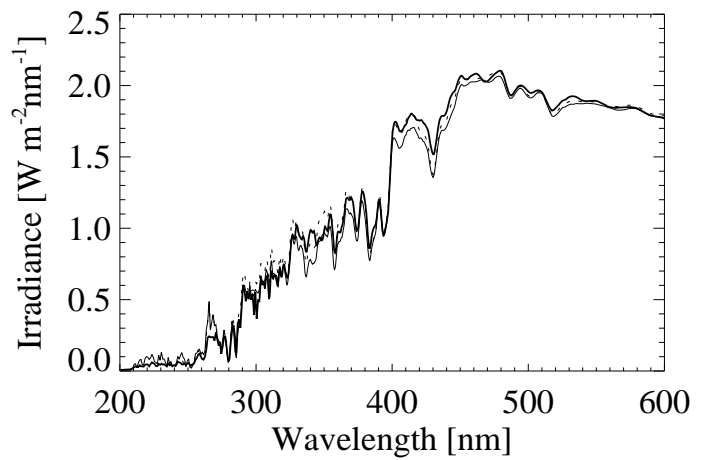}}
\caption{The irradiance calculated by COSI in LTE (thin solid line) and the ATLAS 12 (dashed line), using K91 atmosphere structure and abundances. The spectrum observed by SORCE is given by the thick solid line. All spectra are convolved with the instrument profile of SORCE.}
\label{fig:LTE_SIM}
\end{figure}

\begin{figure*}
\centering
\includegraphics{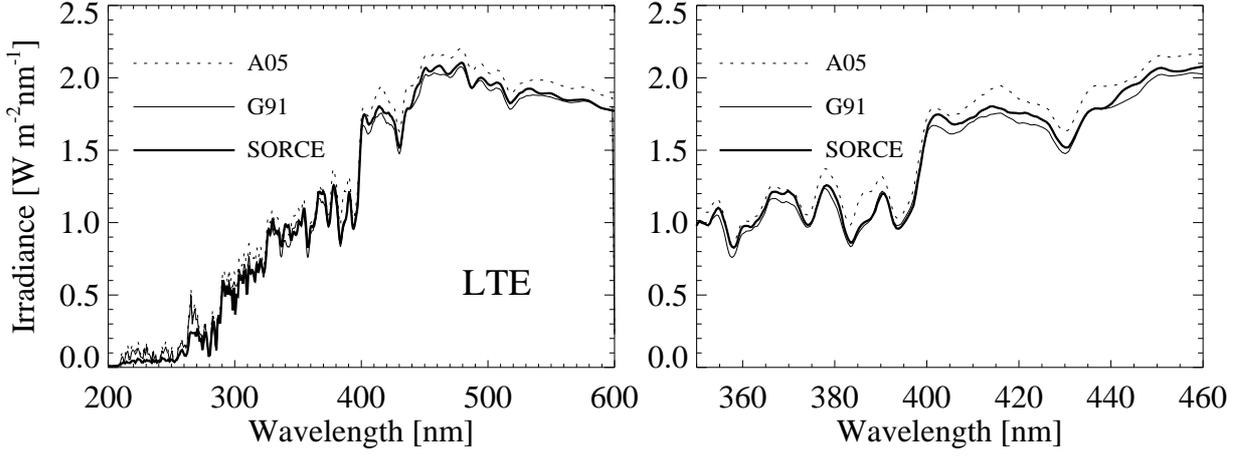}
\caption{Spectra calculated by COSI in LTE using the FALC99 atmosphere structure and A05 and G91 abundances, vs. SORCE measurements. All spectra are convolved with the instrument profile of SORCE.}
\label{fig:LTE_G_A}
\end{figure*}

In Fig.~\ref{fig:LTE_SIM} we compare the spectra calculated by both codes with the SORCE measurements.  One can see that both codes predict the correct value of the  continuum in the red part of the visible spectrum (600 nm) but fail to reproduce the molecular bands and UV part of the spectrum.
The LTE spectra calculated with the FALC99 and two sets of abundances by \citet{grevesseanders1991} (hereafter G91) and by \citet{asplundetal2005}   (hereafter A05) are presented in Fig.~\ref{fig:LTE_G_A}. The calculations with the G91 abundances predict the correct value of the visible and IR continua. Furthermore, they give quite a good fit of molecular bands (see right panel of the picture where the calculation of CN violet system and CN G band are shown in more detail).  However the calculated irradiance in the UV part is again much higher than observed for both abundance sets.

The metal abundances by A05 are significantly lower (up to about 2 times) than in G91.  The lower metal abundance leads to a significantly lower electron density,  consequently to a lower concentration of negative hydrogen ion.  The latter is the main source of the continuum opacity in the solar atmosphere \citep[cf.][p. 102]{mihalas1978}. Therefore the application of the A05 abundances results  in a lower continuum opacity for the visible and IR wavelengths and as a consequence in a higher irradiance. This effect is especially prominent at longer wavelengths (from about 400 nm) where the number of strong lines is relatively low and the opacity of negative hydrogen dominates  all other sources of opacity.
In turn, the lower metal abundance given by A05 leads to lower line opacities and photoionization opacities in the UV. As a result, the A05 abundances also lead to a  higher irradiance in the UV than the G91 abundances. The FALC99 model has so far mainly been used with G91 abundances.

\begin{figure*}
\centering
\includegraphics{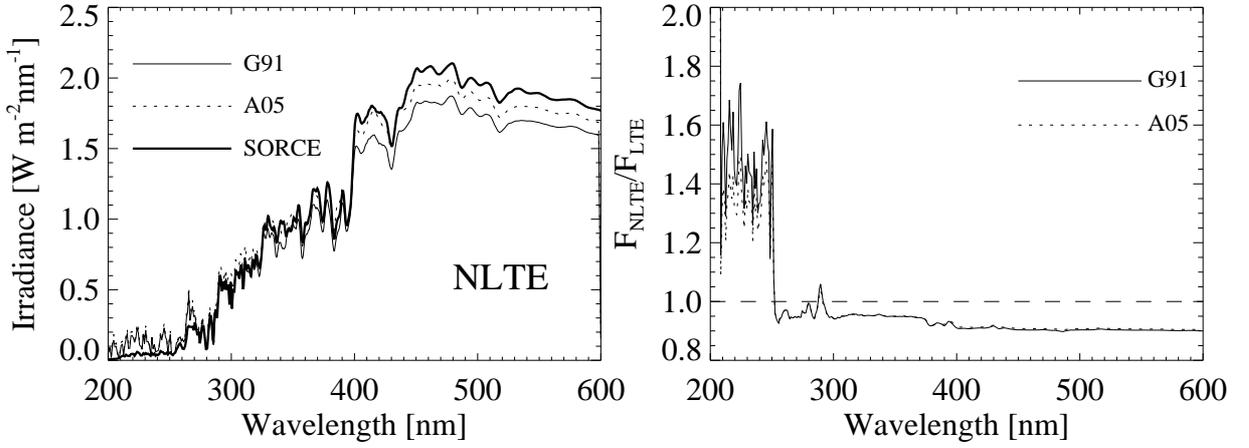}
\caption{Left panel: as the left panel in Fig.~\ref{fig:LTE_G_A} but calculated in NLTE. Right panel: The ratio between NLTE and LTE emergent flux.}
\label{fig:NLTE_G_A}
\end{figure*}

\begin{figure*}
\centering
\includegraphics{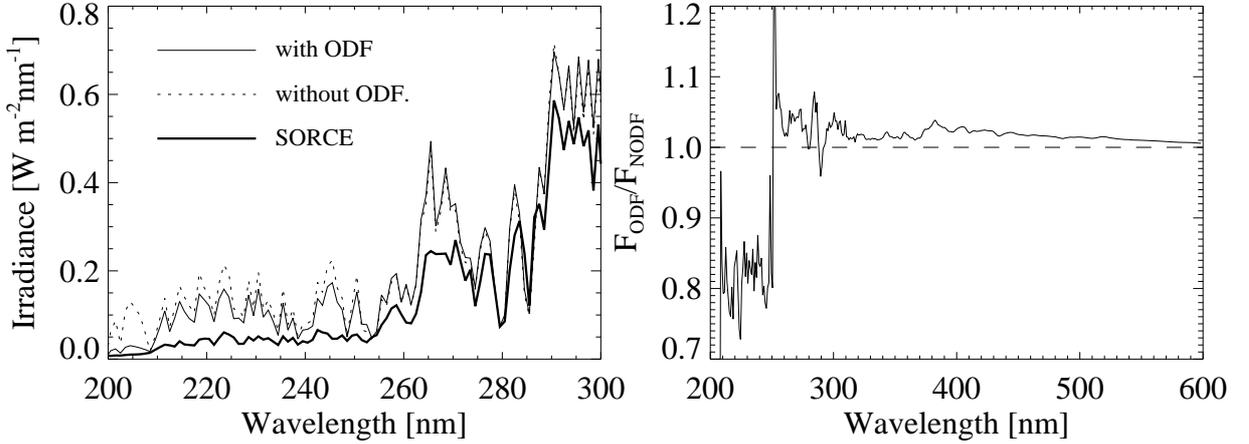}
\caption{Left panel: NLTE calculations with and without iterated ODF, compared to the solar irradiance as observed by SORCE. Right panel: Ratio between the emergent flux from NLTE calculated with and without iterated ODF. All calculations are done with the FALC atmosphere model and G91 abundances.}
\label{fig:ODF}
\end{figure*}

\section{NLTE calculations}\label{sec:NLTEproblems}
Most of the solar radiation emerges from regions  of the atmosphere with a steep temperature gradient. Therefore a significant part of the escaping photons is not  in the thermodynamic  equilibrium  with the surrounding medium. This implies that a self-consistent radiative transfer model has to account for NLTE effects. It is well-known that these effects are extremely important for the strong lines and the UV radiation.  In this section we present NLTE calculations of the solar spectrum with COSI and show that the NLTE effects are also important for the overall energy  distribution in the solar spectrum, including the visible and infrared wavelength ranges. We want to emphasize here that the NLTE calculations presented in this paper were performed using temperature and pressure profiles of the solar atmosphere obtained by \citet{fontenlaetal1999}. 
These profiles in turn depend on assumptions, so all effects presented in this paper are relative to the assumed model of the solar atmosphere (see also the discussion at the end of Sect.~\ref{subsubsec:compdissc}).

In the left panel of Fig.~\ref{fig:NLTE_G_A}, the NLTE calculations of the solar spectrum with the atmosphere structure FALC99 and the abundance sets A05 and G91  are compared to the observed solar spectrum.  The right panel of Fig.~\ref{fig:NLTE_G_A}  shows the ratios of the NLTE to the LTE calculations. The  most prominent overall NLTE effects are the strong increase in the irradiance in the UV and mild decrease of the irradiance in the  red part of the spectrum.  The first effect is discussed in detail in Sect.~\ref{subsec:NLTEUV}, and the second one in Sect.~\ref{subsec:NLTEcont}.

\subsection{Calculating of the UV radiation}\label{subsec:NLTEUV}
\subsubsection{NLTE opacity effects}
Independent of the solar atmosphere model and abundances, the LTE calculations overestimate the UV irradiance up to about 300 nm (see Figs. \ref{fig:LTE_SIM} and \ref{fig:LTE_G_A}). This problem is even more severe in the NLTE calculations as they predict  a UV irradiance that is higher than in the case of the LTE calculations  (see left panel of Fig.~\ref{fig:NLTE_G_A}).

The effect of the increase in the far UV flux in the NLTE calculations is actually well known \citep[cf.][]{shorthauschildt2009}.
The UV radiation incident from the higher and hotter parts of the solar atmosphere onto the photosphere causes  a stronger ionization of  iron and other metals relative to the LTE case. This decreases the populations of the neutral atoms and consequently weakens the strength of the corresponding line and photoionization cross section, which are the main sources of the opacity in the far-UV.  Therefore it leads to a decrease in the opacity and consequently an increase in the flux in the wavelengths up to about 250 nm (red threshold of the photoionization from the Mg I 2 level).  In the longer wavelengths the line and the photoionization opacities of the ionized metals become more important so the increase in their populations  results in lower irradiance.  

The effect of the increased UV flux in the NLTE calculations can be decreased significantly by the use of the opacity distribution function (see Fig.~\ref{fig:ODF}), which allows  incorporation of the opacity from the background lines into the solution of statistical equilibrium equations \citep{haberreiter2008}.  This background opacity diminishes the amount of the penetrating UV radiation and decreases the degree of metal ionization. It leads to the relative decrease in the spectral irradiance up to 250 nm and an increase at higher wavelengths (in agreement with the above discussion).  The complicated wavelength dependence of the spectrum alterations caused by the influence of the ODF  can be explained by the large number of the strong metal lines and photoionization thresholds. Overall,  introducing  the ODF causes the redistribution of the solar irradiance over the entire spectrum as it decreases the far-UV part and increases the near-UV and visible. 

Although the concept of the ODF allows the agreement between calculations and observations to be imporved, the ODF based on the Kurucz linelist \citep{haberreiter2008} cannot solve the discrepancy with measurements. The most probable source of the problem is the inaccuracy of the atomic and molecular line list and possible missing continuum opacity. The production of a complete line list is an extremely laborious task. Considerable progress has been reached during past decades thanks to the effort of the OPACITY project  \citep{seatonnew2005} and, in particular, of \citet{kurucz1993} in calculating atomic data. Concerning the problem of the UV blanketing, it is important to keep in mind that about 99\%(!) of the atomic and molecular lines are predicetd theoretically and only 1\% of all lines are observed in the laboratory \citep{kurucz2005}. This leads to potential errors in the total opacity as lines could still be missing while wavelengths and oscillator strengths of the predicted lines could be inaccurate.  An incompleteness of the atomic data is especially significant in the UV where the immense number of lines  
form the UV line haze.

The consistent  account for the line blanketing effect is very important for the overall absolute flux distribution, especially for the UV \citep[cf.][]{collet2005, avrettloeser2008, fontenlaetal2009}. \citet{busaetal2001} used version 2.2 of the MULTI code, which UV opacity were contributed only from several hundred calculated in the NLTE lines,   bound-free and free-free processes. They showed that NLTE calculations without proper account for the line blanketing effect can lead to overestimating of the UV flux by six orders of magnitude in the case of cold metal-rich stars. To compensate for this, they multiplied the continuum opacity by a wavelength-dependent factor and gave an explicit expression for the factor parameterization.  A similar approach is used by \citet{bruls1993} to calculate the NLTE populations of Ni I.
 \citet{shorthauschildt2009} show that the use of different line lists for the line blanketing calculations can lead to significantly different spectral irradiance distributions. They also show that the problem of the excess of the UV flux in the 300-420 nm spectral region  can be solved by increasing of the continuum absorption coefficient.

 \subsubsection{Additional opacities in COSI} \label{subsubsec:addopac}
COSI takes the opacity  from the several million lines into account \citep{haberreiter2008}. However, the significant disagreement  between the calculated and measured solar UV flux as discussed above indicates that the computations still miss an essential part of the opacity. To compensate for this missing opacity,  \citet{haberreiter2008} calculated the UV part of the solar spectrum  with artificially increased Doppler broadening.
This approach proves itself successful, but it cannot be used to fit high-resolution spectra. Therefore, here we have extended the approach by \citet{busaetal2001} and \citet{shorthauschildt2009}  and included additional opacity in our NLTE-ODF scheme, as described below.

COSI overestimates the irradiance, i.e., shows an opacity deficiency in the spectral region between 160 nm and 320 nm. At shorter wavelengths, the opacity is dominated by continuum photoionization, while at longer wavelengths tabulated atomic and molecular lines are sufficient to reproduce the observations. In this wavelength region, we grouped the solar spectrum into 1 nm intervals (corresponding to the resolution of the SOLSTICE/SORCE spectrum). For each interval we empirically determined the coefficient $f_{\rm c}(\lambda)$  by which the continuum opacity has to be multiplied to reproduce the SOLSTICE measurements. The additional emission coefficient was calculated under the assumption that the missing opacity source has thermal structure, so its source function is equal to the Planck function.  Because the correction of the ODF affects the statistical equilibrium of the populations, we iteratively solve  the statistical equilibrium equations and the factors $f_{\rm c}(\lambda)$ using newly updated ODFs until we reached a self-consistent solution. Thus, the $f_{\rm c}(\lambda)$ factors were found iteratively. We stopped the iteration when the deviations to the observed spectrum was smaller than 10\%. In the considered region there were very few wavelength points where the irradiance was underestimated. We did not apply any correction to these points. This underestimation can be explained by the inaccuracy of the line list because it can be caused by the small wavelength inaccuracy of a few strong lines.

\begin{figure}
\resizebox{\hsize}{!}{\includegraphics{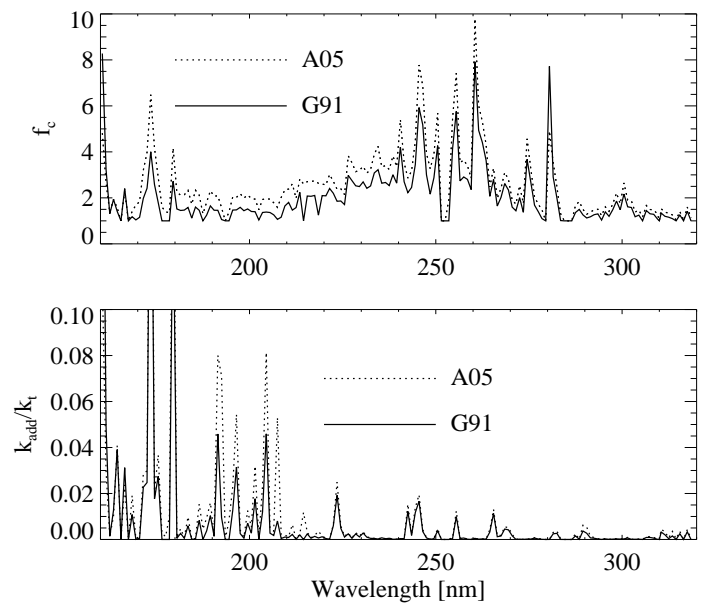}}
\caption{The factor $f_{\rm c}(\lambda)$ (upper panel) and the ratio of the additional and total opacity at the height 200 km (lower panel) for the G91 and A05 abundances.   }
\label{fig:fudge}
\end{figure}

The wavelength dependence of $f_{\rm c}(\lambda)$ is shown in Fig.~\ref{fig:fudge} for the calculations with the FALC99 atmosphere model and G91 and A05 sets of abundances. One can see that $f_{\rm c}(\lambda)$ is very close to $1$ (which means little additional opacity) for wavelengths longer than 280 nm. There are also several prominent peaks that indicate a strong lack of opacity at specific wavelengths. Although different abundances lead to a change in the peak amplitudes, the overall profile of wavelength dependence stays the same for both sets of abundances, G91 and A05, respectively.
The averaged line opacity is higher by several orders of magnitude than the continuum opacity, so the ratio of the additional opacity, which was needed to achieve agreement with the observed solar spectrum, to the total opacity that was already included into COSI is for most wavelength below 5\ \%\ (see lower panel of Fig.~\ref{fig:fudge} ).
Therefore, the correction of the overestimation of the emergent UV flux can be achieved by a moderate modification of the input line list.

Calculations with the G91 abundances lead to a lower continuum than calculations with A05 abundances. Therefore the flux overestimation is stronger in the case of calculations with A05 abundances. Consequently one has to use larger correction factors $f_{\rm c}(\lambda)$ for A05 abundances.
The correction factor strongly depends on the chosen line list. We tested our calculations using the Vienna Atomic Line Database (VALD) \citep{kupkaetal1999, kupkaetal2000} and retrieving the lines with known radiative damping parameter. Because this database contains fewer lines then provided by \citet{kurucz2006}, a larger factor $f_{\rm c}(\lambda)$ is needed to achieve agreement with the observations

\subsection{Calculation of the visible and IR radiation}\label{subsec:NLTEcont}
NLTE effects decrease the visible and near IR irradiance by about 10\ \%\ (see Fig.~\ref{fig:NLTE_G_A}), because they influence the concentration of  electrons and the negative hydrogen ions that determines the continuum opacity, as well as the continuum source function.

\subsubsection{NLTE effects in electron and negative hydrogen concentrations}\label{subsubsec:Hminus}
\begin{figure}
\resizebox{\hsize}{!}{\includegraphics{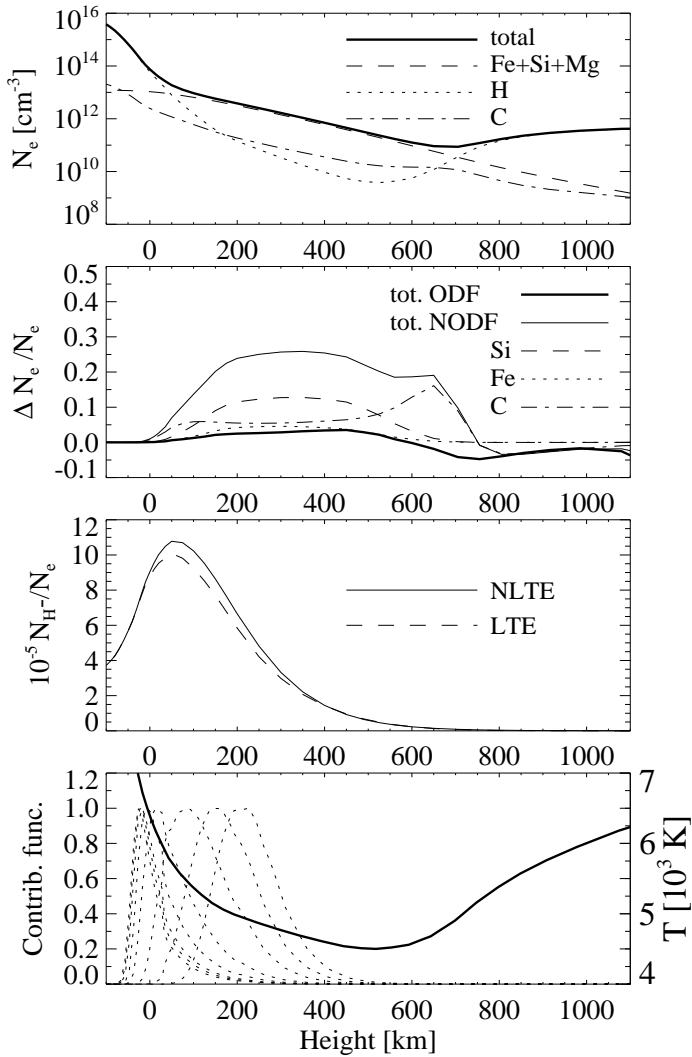}}
\caption{NLTE effects in electron and $ \rm {H^{-}}$ concentrations. Upper panel: depth dependence of the total electron concentration and the electron concentrations resulting from the ionization of elements as indicated.  Second panel:  ratios of the change of the electron concentrations due to the NLTE effects and total electron concentration calculated in LTE. Plotted are total changes with and without applying ODF and resulting from the ionization of particular elements change without ODF. Third panel: depth dependence of the ratio of negative hydrogen and electron concentration  calculated in NLTE and LTE. Lower panel: Contribution functions to the intensity for the $5000$  ${\AA}$ (dotted lines) continuum radiation and temperature on depth dependence (solid curve) for the FALC99 atmosphere model. The contribution functions are plotted for eight values of the cosine of 
the angle between the propagation direction of radiation and the local solar radius $\mu$. From right to left: $\mu=$
0.05 (almost solar limb), 0.1, 0.2, 0.4, 0.6, 0.8, 0.9, 1.0 (solar disk center). All calculations are carried out for the FALC99 atmosphere model and A05 abundances}
\label{fig:elconc}
\end{figure}

Departures from LTE in the electron and hydrogen negative ion concentrations are presented in Fig.~\ref{fig:elconc}  for the calculations with FALC99 atmosphere structure and A05 abundances. The upper panel illustrates the main electron sources in the solar atmosphere. In the lower part of the photosphere and in the chromosphere, hydrogen is ionized and provides most of the electrons; however, in the photosphere, which is responsible for the formation of the visible and near-IR irradiance, the temperature is too low for ionizing hydrogen.  Therefore, here the electrons are mainly provided by the metals with low ionization potential, such as iron, silicon, and magnesium.

As one can see from the second upper panel of Fig.~\ref{fig:elconc},  the ionization of silicon and iron is strongly affected by the NLTE effects. This is basically the same effect of NLTE  ``overionization'' as already discussed in Sect.~\ref{subsec:NLTEUV}, where the main emphasis was on the deviations in the concentration of the neutral atoms. Applying of the ODF significantly decreases the NLTE effects on the electron concentration as it strongly increases the UV opacity. In the last part of Fig.~\ref{fig:elconc}, the contribution functions for the continuum radiation are plotted for different $\mu$-positions.  The contribution functions show where the emergent intensity originates \citep[see, e.g.][p.\,151]{gray1992}, so one can see that, although the continuum is mainly formed in the lower part of the photosphere where the NLTE effects in the electron concentration are not so prominent, they still can be important,  especially for the radiation emitted close to the solar limb.

The photoionization cross section of negative hydrogen has its threshold at about 16\,500 $\AA$, and it reaches its maximum at about 8500 $\AA$ \citep[cf.][p. 102]{mihalas1978}. Although the photosphere is optically thick for the UV  and radiation in the strong atomic or molecular lines, the visible  and IR photons can easily escape, even from the lower part of the photosphere. (The optical depth of the zero point in the continuum radiation is about one, depending on the wavelengths.) This means that all over the photosphere there is a lack of photons that are able to photoionize the negative hydrogen. Thus, in the NLTE case, the radiative destruction rate of H$^-$ is decreased, which in turn increases the concentration of the negative hydrogen ion compared to the LTE equilibrium. This is illustrated in the third panel of Fig.~\ref{fig:elconc}. The effect is especially strong in the layers where the continuum radiation is formed (see the contribution functions in the lower panel of Fig.~\ref{fig:elconc}). Moreover, in contrast to effects on the electron concentration, applying  the ODF  does not decrease the NLTE deviations of the negative hydrogen concentration  since UV radiation does not make any significant contribution to its ionizations. The effect of  negative hydrogen underionization was firstly pointed out by \citet{vernazzaetal1981}, who found that photospheric departure coefficients of the negative hydrogen (ratios between NLTE and LTE concentrations) are greater than one.

Thus, compared to the LTE case, NLTE ionization of hydrogen and metals result in higher electron concentration. The NLTE ionization of negative hydrogen also leads to the increase in its concentrations. This enlarges the continuum opacity so that continuum radiation comes from higher and cooler layers of the photosphere. In addition, there is also a decrease in the NLTE continuum source function, which contributes to the decrease in the continuum emission.

The effects described above significantly decrease the level of the emergent flux in the visible and near IR. Although the NLTE treatment is physically more consistent than LTE, it leads to a severe deviation from the observed solar spectrum (see Fig.~\ref{fig:NLTE_G_A}).

\subsubsection{Collisions with negative hydrogen}

One possible cause of the inability of the NLTE calculations to reproduce the measurements  is the inadequate treatment of collisional rates for the negative hydrogen ion. The present version of COSI accounts for the electron detachment through collisions with electrons and neutral hydrogen, as well as charge neutralization with protons as given by  \citet{lambertpagel1968}.  We do not treat collisions of H$^-$ with heavy elements, which in principle could lead to a strong underestimation of the collisional rates. We found that increasing the  collisional rates involving H$^-$ by a factor of ten solves the disagreement between the measured and calculated continuum flux for the models using A05 abundances. 
However, when using increased collisional rates, the molecular bands (especially CH G band) appear to be weaker than the observed ones. To make the molecular lines consistent with observations we changed the A05 carbon and nitrogen abundances and used them as given by G91. This allows us to reach a very good agreement with measurements (see Sect.~\ref{subsubsec:compdissc}).

Let us emphasize that an increase in the collisional rates for H$^-$ cannot  help  adjust the NLTE calculations performed with G91 abundances, because the considered increase in collisions can only enforce  an LTE population ratio between negative hydrogen and H I 1 level. The latter is, however, in  an NLTE regime  mainly thanks to the NLTE ``overionization''. As the calculations with G91 abundances give  good agreement in a purely LTE regime (see Fig.~\ref{fig:LTE_G_A}), the considered increase in the collisional rates for H$^-$ is not enough to enforce full LTE agreement. An NLTE treatment of the solar atmosphere yields population numbers of negative hydrogen that strongly deviate from LTE.

\subsubsection{Change in abundances}
\begin{figure*}
\centering
\includegraphics{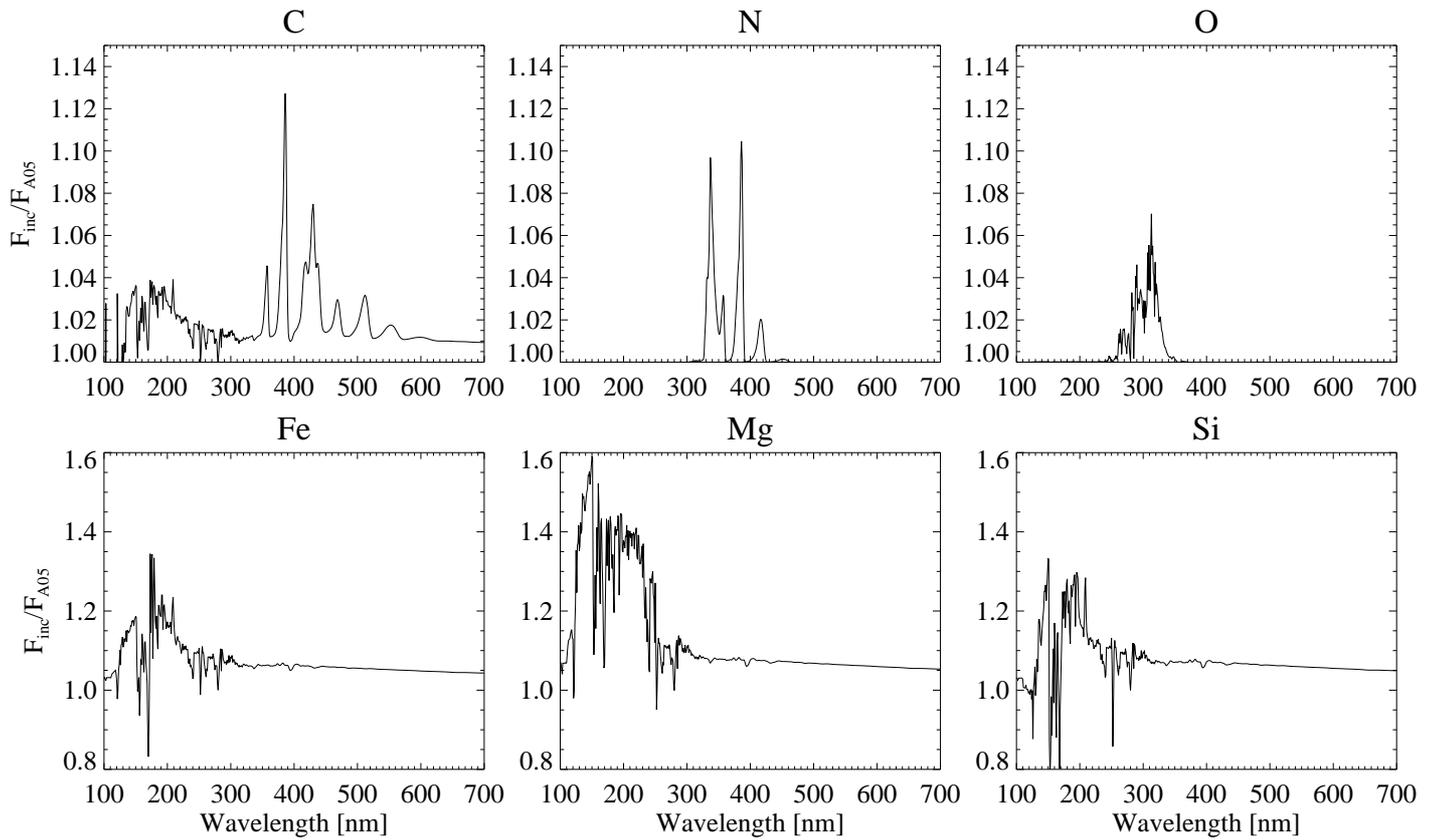}
\caption{ Ratios of the emergent fluxes  calculated with A05 abundances to the fluxes calculated with the abundances of the indicated element increased by a factor of two. } 
\label{fig:ab}
\end{figure*}

The effect of using of different abundances on the emergent synthetic spectrum is illustrated in Fig.~\ref{fig:ab}. One can see that the abundance change of elements like iron, magnesium, and silicium strongly affects the UV radiation (mainly due to the photoionization opacities), but also the visible and infrared continuum, because these elements are the main donors of the photospheric electrons. In contrast, a change in nitrogen and oxygen abundances has no effect on the continuum and only affects the spectrum via the molecular bands. The carbon abundance slightly influences the continuum, but the main effect of carbon in the visible spectral range comes from molecular systems. The CN violet system at about 380 nm and CH G band at about 430 nm are the most prominent features.
Thus, the carbon and nitrogen abundance changes from the A05 to G91 values almost do not affect the continuum level (see Fig.~\ref{fig:ab}) and lead to  good agreement with the measurements (see Fig.~\ref{fig:good_git}). Therefore, we can conclude that A05 abundances should be given preference for calculating the NLTE continuum, except for carbon and nitrogen, for which the G91 abundances lead to better reproduction of the molecular bands. The favored combination of abundances leads to good overall agreement with the observed solar spectrum.

\subsection{Comparison with measurements and discussion}\label{subsubsec:compdissc}
\begin{figure*}
\centering
\includegraphics{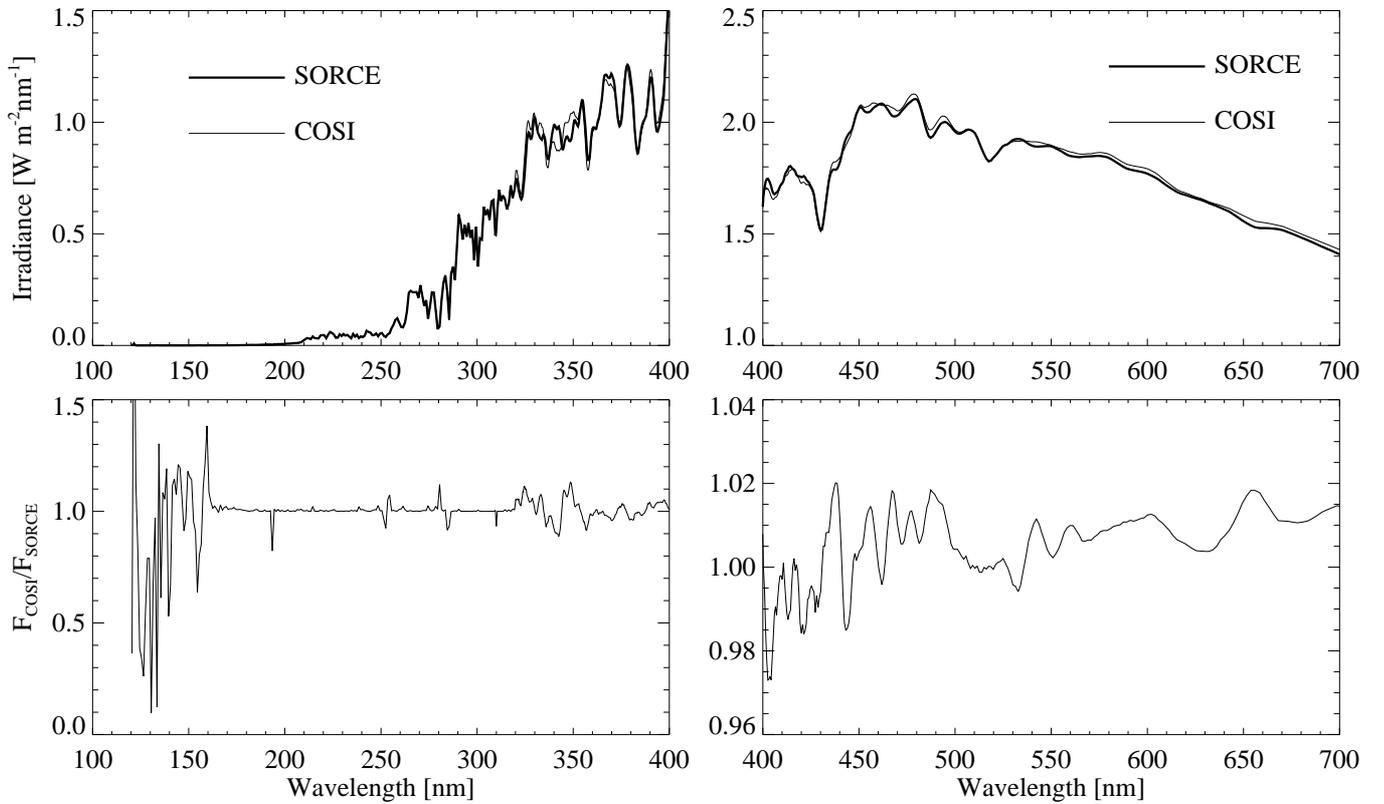}
\caption{Upper panel: SORCE observations  vs. calculations  with the FALC atmosphere model and described in the text abundances.  Lower panel: ratio between the calculations and measurements.}
\label{fig:good_git}
\end{figure*}

In Fig.~\ref{fig:good_git} we present the solar spectrum calculated with the FALC99 atmosphere model compared with the SORCE measurements. The agreement in the 160-320 nm spectral region is automatically good (see Sect.~\ref{subsubsec:addopac}). At shorter wavelengths, the observed flux is determined by several strong emission lines, which are treated in LTE in the current version of COSI. This leads to a strong underestimation of the irradiance. In a future version of COSI,  these lines will be treated in NLTE. 
The Ly $\alpha $ line has already been computed in NLTE, and we obtain an overestimation of its flux. In a forthcoming paper we will investigate it in more detail  \citep{schoelletal2009}.

\begin{figure*}
\centering
\includegraphics{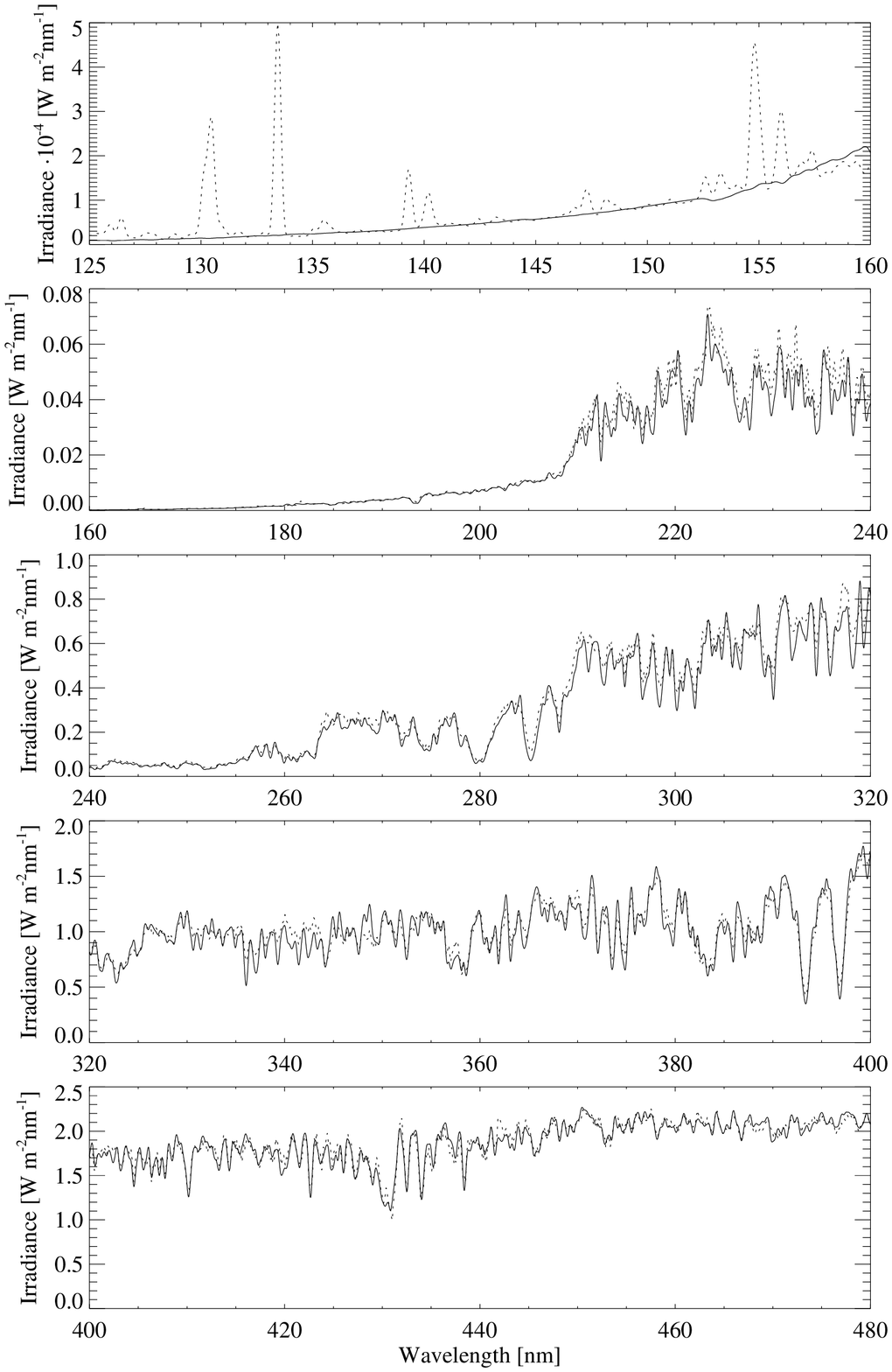}
\caption{The spectra calculated with COSI (solid line) vs.  SOLSPEC measuremets (dotted line). }
\label{fig:SOLSPEC1}
\end{figure*}

\begin{figure*}
\centering
\includegraphics{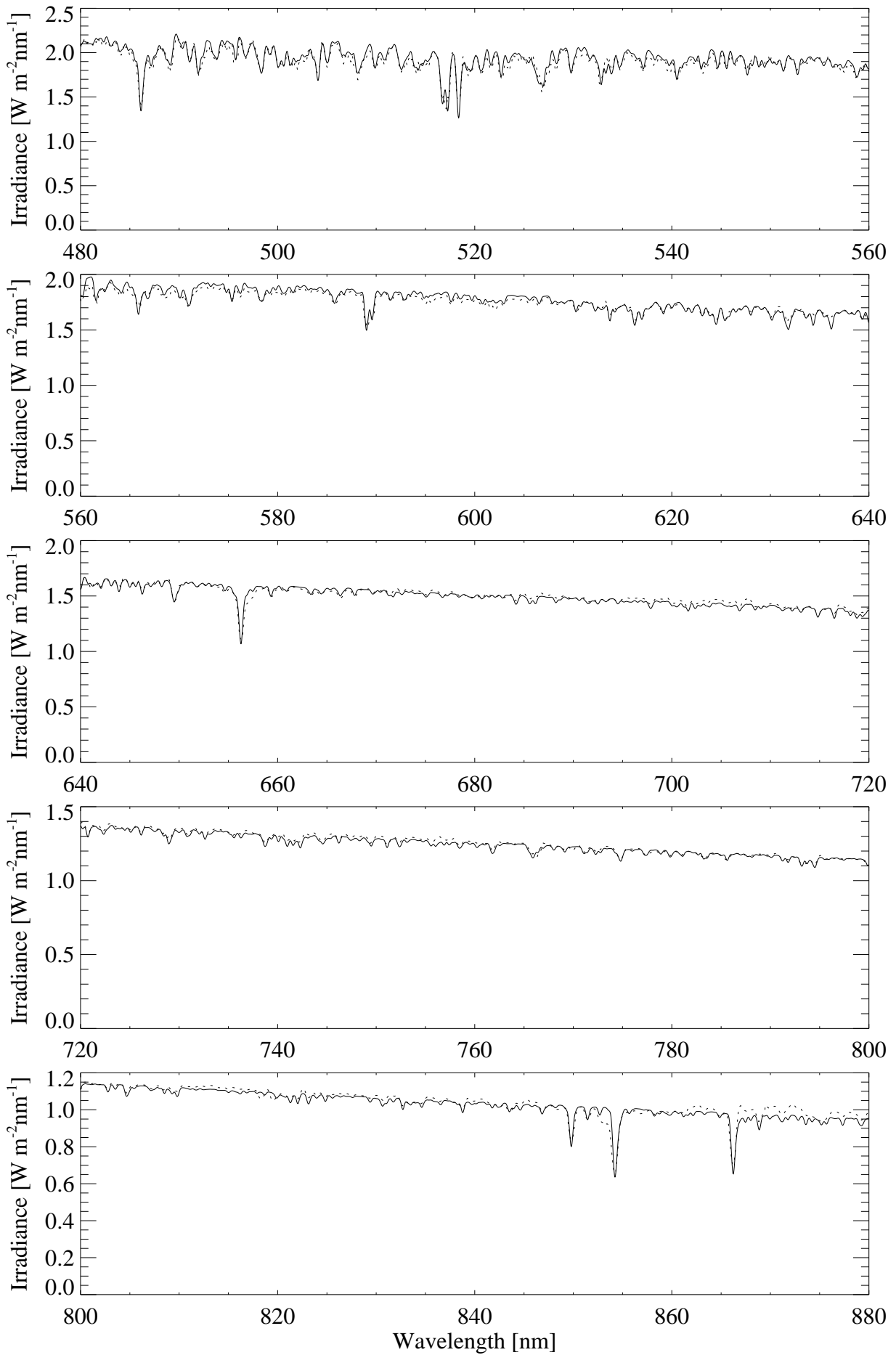}
\caption{As Fig.~\ref{fig:SOLSPEC1}. Continuation. }
\label{fig:SOLSPEC2}
\end{figure*}

 The detailed comparison of our synthetic spectrum with SOLSPEC measurements is given in Figs.~\ref{fig:SOLSPEC1} and \ref{fig:SOLSPEC2} in the Online Material. One can see that starting from 160 nm  two spectra are in remarkable agreement with each other. Although several strong emission lines are currently not properly  calculated   in shorter wavelengths, the continuum level there is also consistent with SOLSPEC measurements. This can only be achieved with NLTE calculations.

Thus the assumption of enhanced negative hydrogen collisions and the use of the combined G91 and A05 abundance set provide us with the possibility of successfully  modeling the entire solar spectrum. 
Another possible cause of the problem that synthetic NLTE models have difficulty fitting the observed solar spectrum is that the FALC99 atmosphere model was developed to fit the visible and IR continuum  with the radiative transfer code PANDORA  
\citep[see][]{fontenlaetal1999}, which uses a different approximate approach to treat the NLTE effects, different sets of the atomic and molecular data as well as different sets of abundances. The \citet{fontenlaetal1999} atmosphere structures have been used successfully in various applications  \citep[cf.][]{penzaetal2004a, penzaetal2004b,vitasvince2005} in which, however, the continuum flux was calculated in LTE.
The quest to construct a self-consistent atmosphere model that fully accounts for all NLTE effects will be addressed in a future investigation.

\begin{figure*}
\centering
\includegraphics{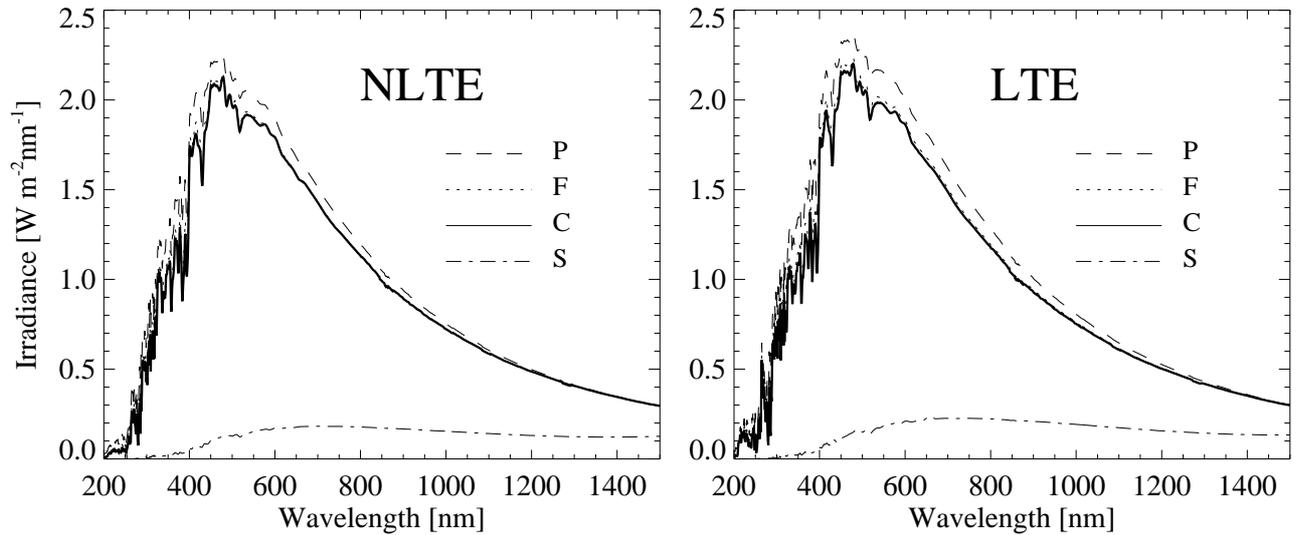}
\caption{NLTE (left panel) and LTE (right panel) calculations of the solar irradiance from the quiet Sun, active network, plage,  and sunspot.}
\label{fig:4models}
\end{figure*}

\begin{figure*}
\includegraphics{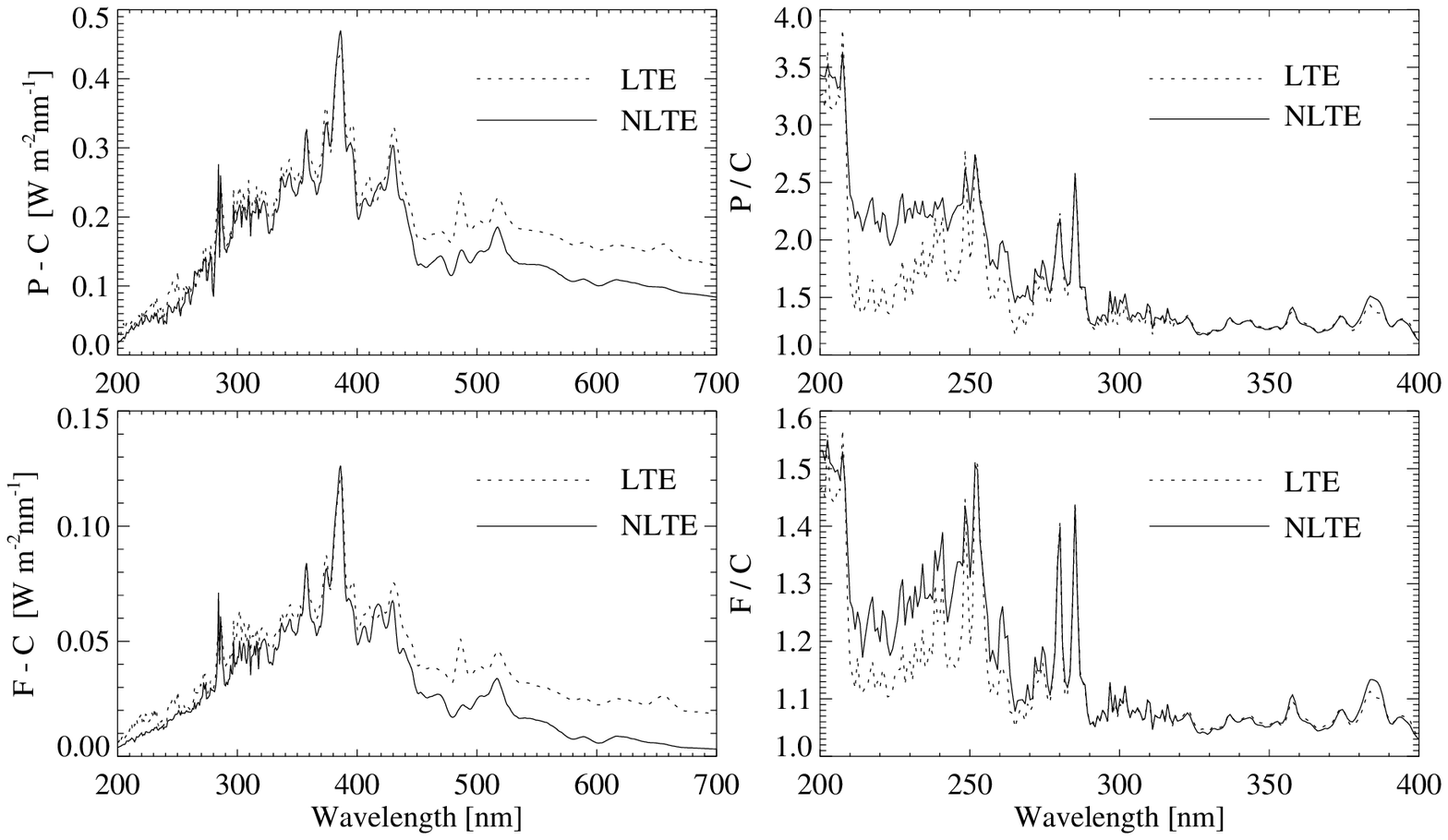}
\caption{The flux differences between plage and quiet Sun (upper panel), and bright network and quiet Sun (lower panel) calculated in NLTE and LTE.}
\label{fig:contrast}
\end{figure*}

\section{Irradiance variations}\label{sec:active}
It is generally accepted that most of the solar variability is introduced by  the competition between the solar flux decrease due to the dark sunspots and a flux increase due to the bright magnetic network and plage \citep[e.g.][]{willsonhudson1991, krivovaetal2003}. It is therefore important to consistently calculate the solar irradiance from these active components of the solar atmosphere.    

For our calculations we used the \citet{fontenlaetal1999} atmosphere model FALC99 for the quiet Sun, model FALF99 for the bright network, model FALP99 for plage, and model FALS99 for the sunspot. Both NLTE and LTE spectra of the quiet Sun, active network, plage, and sunspot are presented in Fig.~\ref{fig:4models}. The bright network spectrum is hardly distinguishable from the quiet Sun spectra. The NLTE spectra produce a lower visible and UV flux and, accordingly,  lower total solar irradiance (TSI). This effect for the quiet Sun irradiance was already discussed in Sect.~\ref{subsec:NLTEcont}. 

Another interesting detail is the decrease in the contrast between the quiet Sun, plage, and bright network in the NLTE calculations. The main reason for this is that the NLTE effects that decrease the continuum source function are stronger in the hotter FALP model as it corresponds to the lower particle density (and consequently lower collisional rates). 

Figure \ref{fig:contrast} shows the flux differences between the active components of the solar atmosphere and the quiet Sun calculated in LTE with A05 abundances, and NLTE with enhanced collisions and combined A05 and G91 abundances (see Sect.~\ref{sec:NLTEproblems}). The most prominent peak at about 3890 $\AA$ corresponds to the CN violet system. Although the differences in the CH G band around 4300 $\AA$  are also clearly visible, they are less pronounced than in the CN violet system.  This is caused mainly by the differences in the CH and CN dissociation energies (3.465 eV and 7.76 eV), so the CN concentration is more sensitive to temperature variations (see Sect.~\ref{subsec:chem}). Furthermore, the contrast between the active components and the quiet Sun is significantly decreased in the NLTE calculations. As a consequence, the NLTE calculations reduce the solar variability in the visible and IR, shifting it to the UV.

The differences between the spectral fluxes for active components and for the quiet Sun integrated from 900 $\AA$ to 40\,000 $\AA$  are presented in Table~\ref{table:irrchange} for the LTE (A05 and G91 abundances), for the NLTE with G91 abundances and  NLTE with enhanced collisions and combined A05 and G91 abundances (NLTE comb.). We conclude that the NLTE calculations (both with and without ODF) significantly reduce the variations in the TSI.

\begin{table}
\caption{TSI differences with FALC model per 1 \% area (in $W/m^2$).} 
\label{table:irrchange}
\centering 
\begin{tabular}{ l  | c  c  c } 
\hline
\hline
      & FALF &   FALP & FALS\\ 
\hline 
  LTE G91                        &    0.28   &  1.48      &  -10.7 \\
   LTE A05                         &      0.25   & 1.27   & -10.82  \\
  NLTE G91                     &       0.18 &  0.76  &  -11.02  \\
  NLTE comb.   &    0.15 & 0.99   &  -11.05  \\
\hline 
\end{tabular}
\end{table}

 \section{Conclusions}\label{sec:conclusions}
We have presented a further development of the radiative transfer code COSI.  The code accounts for the NLTE effects in several hundred lines, while the NLTE effects in the several million other lines are indirectly included via iterated opacity distribution function. The radiative transfer is solved in spherical symmetry.
The main conclusions can be summarized as follows.
\begin{itemize}
          \item The inclusion of the molecular lines in COSI allowed us to reach good agreement with the SORCE observations in the main molecular bands (especially in the CN violet system and CH G band). It has also solved the previous  discrepancies  between the LTE calculations with the  COSI code and ATLAS 12  calculations. We showed that their strong temperature sensitivity allows molecular lines to significantly contribute to the solar irradiance variability. 
          \item We introduced additional  opacities into the opacity distribution function. It allowed us to solve the well-known problem of overestimating the UV flux in the synthetic spectrum.  We should emphasize, however, that the magnitude and behavior of the additional opacity strongly depend on the applied model. 
	\item We have shown that the concentration of negative hydrogen is strongly affected by NLTE effects as explained in Sect.~\ref{subsubsec:Hminus}. It decreases the level of the visible and infrared continuum and leads to a discrepancy with the measured level if the calculations are done with the current models of the quiet Sun atmosphere. 
	\item We presented  calculations of the total and spectral solar irradiance changes due to the presence of the active regions and showed that NLTE effects can strongly affect both of these values.

\end{itemize}

\begin{acknowledgements}

These investigations have benefited from stimulating discussions at meetings organized at the ISSI in Bern, Switzerland.
The research leading to this paper received funding from the
European Community's Seventh Framework Program (FP7/2007-2013) under
grant agreement N 218816 (SOTERIA project, www.soteria-space.eu). We thank J. Fontenla for supplying the SRPM database.

\end{acknowledgements}


\bibliographystyle{aa}
\bibliography{shapiro}


\Online

\end{document}